\documentclass[conference]{IEEEtran}
\IEEEoverridecommandlockouts
\usepackage{cite}
\usepackage{amsmath,amssymb,amsfonts}
\usepackage{algorithmic}
\usepackage{graphicx}
\usepackage{textcomp}
\usepackage{xcolor}
\usepackage{url}
\def\BibTeX{{\rm B\kern-.05em{\sc i\kern-.025em b}\kern-.08em
    T\kern-.1667em\lower.7ex\hbox{E}\kern-.125emX}}

\begin{document}

\title{QKD-KEM: Hybrid QKD Integration into TLS with OpenSSL Providers}
%El título está centrado en el trabajo de colaboración conjunto... pero se puede ampliar/cambiar a lo que consideréis

\author{
\IEEEauthorblockN{Javier Blanco-Romero}
\IEEEauthorblockA{\textit{Department of Telematic Engineering} \\
\textit{Universidad Carlos III de Madrid}\\
Leganés, Madrid, Spain \\
frblanco@pa.uc3m.es}
\and
\IEEEauthorblockN{Pedro Otero García}
\IEEEauthorblockA{\textit{atlanTTic Research Center (IC Lab)} \\
\textit{University of Vigo}\\
Spain \\
pedro.otero@det.uvigo.es}
\and
\IEEEauthorblockN{Daniel Sobral-Blanco}
\IEEEauthorblockA{\textit{Department of Telematic Engineering} \\
\textit{Universidad Carlos III de Madrid}\\
Leganés, Madrid, Spain \\
dsobral@pa.uc3m.es}
\and
\IEEEauthorblockN{Florina Almenares Mendoza}
\IEEEauthorblockA{\textit{Department of Telematic Engineering} \\
\textit{Universidad Carlos III de Madrid}\\
Leganés, Madrid, Spain \\
florina@it.uc3m.es}
\and
\IEEEauthorblockN{Ana Fernández Vilas}
\IEEEauthorblockA{\textit{atlanTTic Research Center (IC Lab)} \\
\textit{University of Vigo (Spain)}\\
avilas@det.uvigo.es}
\and
\IEEEauthorblockN{Rebeca P. Díaz-Redondo}
\IEEEauthorblockA{\textit{atlanTTic Research Center (IC Lab)} \\
\textit{University of Vigo (Spain)}\\
rebeca@det.uvigo.es}
}

\maketitle

\begin{abstract}
%\raggedright
Quantum Key Distribution (QKD) promises information-theoretic security, yet integrating QKD into existing protocols like TLS remains challenging due to its fundamentally different operational model. In this paper, we propose a hybrid QKD-KEM protocol with two distinct integration approaches: a client-initiated flow compatible with both ETSI 004 and 014 specifications, and a server-initiated flow similar to existing work but limited to stateless ETSI 014 APIs. Unlike previous implementations, our work specifically addresses the integration of stateful QKD key exchange protocols (ETSI 004) which is essential for production QKD networks but has remained largely unexplored. By adapting OpenSSL's provider infrastructure to accommodate QKD's pre-distributed key model, we maintain compatibility with current TLS implementations while offering dual layers of security. Performance evaluations demonstrate the feasibility of our hybrid scheme with acceptable overhead, showing that robust security against quantum threats is achievable while addressing the unique requirements of different QKD API specifications.
\end{abstract}


\section{Introduction}

The Transport Layer Security (TLS) protocol is the cornerstone of secure Internet communication, with TLS 1.3~\cite{rescorla2018transport} introducing significant security and performance improvements. However, quantum computing threatens classical cryptographic primitives, necessitating quantum-resistant alternatives.

Quantum Key Distribution offers information-theoretic security through quantum mechanical principles~\cite{gisin2002quantum}, though cybersecurity agencies~\cite{QKD2024} note its limited maturity in practical deployments. We propose hybrid approaches combining QKD with post-quantum cryptography (PQC)~\cite{bernstein2017post} to address these limitations. Integration challenges arise from fundamental differences: traditional key exchanges derive secrets from public values, while QKD uses pre-established keys distributed via quantum channels.

Recent standardization efforts related to TLS 1.3 have established a framework for hybrid key exchange that combines traditional and post-quantum cryptography~\cite{stebila2020hybrid}. Their approach demonstrates how to achieve security through the concatenation of shared secrets from different key exchange methods, maintaining protection as long as at least one component remains unbroken. This design principle is relevant for QKD deployments where extra security is needed: while QKD provides information-theoretic security, supplementing it with post-quantum cryptography can provide additional protection against implementation vulnerabilities or operational compromises in the QKD system.

Our work explores QKD integration into TLS using OpenSSL's provider infrastructure, encapsulating both PQC shared secrets and QKD key identifiers into a single KEM operation. Unlike previous implementations, we specifically address the integration of stateful QKD protocols through ETSI 004 API. Our implementation maintains compatibility with existing TLS while providing dual security layers, demonstrating a viable pathway for quantum-safe TLS with acceptable performance overhead.

The remainder of this paper is organized as follows. Section~\ref{sec:soa} provides an overview of the integration of QKD with secure communication protocols, discussing the related work. Section~\ref{sec:proposal} presents the proposed approach, detailing its design, implementation, and key components. In Section~\ref{sec:evaluation}, we evaluate the performance and security of our method through preliminary experiments and show the results. Finally, Section~\ref{sec:conclusions} concludes the paper, summarizing our findings and outlining future research directions.

\section{Related Work}
\label{sec:soa}

The integration of QKD with cryptographic protocols has evolved from early IPSec-focused approaches to recent implementations in TLS. Sfaxi et al. \cite{sfaxi2005using} proposed SeQKEIP, introducing a preliminary QKD phase for IPSec, while Berzanskis et al. \cite{berzanskis2009method} presented a hybrid method combining QKD and classical keys through XOR operations. Aguado et al. \cite{aguado2017hybrid} extended Diffie-Hellman by including QKD key identifiers in both SSH and SSL/TLS.

Rijsman et al. \cite{rijsman2019openssl} demonstrated QKD integration in OpenSSL through its engine interface without altering the core code. Later, Dowling et al. \cite{dowling2020many} and Huang et al. \cite{huang2020practical} proposed hybrid Authenticated Key Exchange protocols that combine classical, post-quantum, and QKD techniques, enhancing overall security even if one component is compromised. Dervisevic et al. \cite{dervisevic2021overview} provided a comprehensive survey of these approaches, emphasizing challenges such as key synchronization and rekeying rates.

Recent research has shifted focus toward TLS integration. Kozlovičs et al. \cite{kozlovivcs2023quantum} introduced a ``virtual" KEM that interfaces with a QKD infrastructure through modified key exchange flows, while Rubio-García et al. \cite{garcia2023enhancing} proposed an experimental quantum-resistant architecture that allocates QKD for fiber links and PQC for wireless connections. In a triple-hybrid TLS 1.3 implementation, Rubio Garcia et al. \cite{garcia2023quantum} concatenated ECDH, CRYSTALS-Kyber, and QKD-derived secrets using a QKD key exchange approach where the client initiates the process by sending a QKD acknowledgment in the Client Hello message, and the server responds with the key ID in the Server Hello message. Further integration with OpenSSL 3.2.0, using classical cryptography and QKD, is presented in \cite{garcia2024integrating}. Rencis et al. \cite{rencis2024hybrid} proposed a QKD-as-a-Service framework employing different cryptographic methods for various network segments. Similarly, Hoque et al. \cite{hoque2024exploring} combined PQC and QKD in mobile networks, although they used them separately in different network segments.

Standardization efforts continue through the draft ITU-T Recommendation \cite{itu2024qkdtls} for QKD-TLS integration. Alia et al. \cite{alia2024100} demonstrated high-speed QKD-IPSec implementation with sequential hybrid key exchange. Despite these advances, a significant gap remains: integrating stateful QKD key exchange protocols like ETSI 004 into TLS frameworks. Our work addresses this by adapting KEM abstraction to accommodate both stateful (ETSI 004) and stateless (ETSI 014) QKD interfaces.

%\section{methodology}
\section{QKD-KEM Proposal}
\label{sec:proposal}

We propose a hybrid QKD-KEM protocol that integrates Quantum Key Distribution with PQC into TLS via OpenSSL's provider infrastructure. Our approach encapsulates both the PQC shared secret and a QKD key identifier within a single key encapsulation mechanism, enabling use in existing TLS implementations.

Our implementation is publicly available in the following repositories:
\begin{itemize}
    \item QKD-KEM Provider \cite{qursa2024qkd}: OpenSSL 3.0 provider implementing hybrid key encapsulation mechanisms combining QKD with post-quantum cryptography
    \item QKD ETSI API \cite{blanco2024qkd}: C wrapper library for ETSI QKD 004 and 014 API specifications
    \item Benchmarking Suite \cite{qursa2024bench}: Performance evaluations for hybrid QKD-KEM operations
\end{itemize}

Traditional key exchanges derive secrets from exchanged public values, whereas QKD pre-establishes keys over quantum channels and uses key identifiers during classical communication. This mismatch poses challenges because OpenSSL 3.x has two distinct provider interfaces for key establishment: the KEYEXCH interface~\cite{openssl2024keyexch} for protocols like Diffie-Hellman where both parties contribute to the shared secret calculation, and the KEM interface~\cite{openssl2024kem} for encapsulation-based methods where one party generates and encapsulates a secret for the other. Neither naturally accommodates QKD's pre-distributed key model, which operates fundamentally differently from both approaches. Similar to how Rijsman et al.~\cite{rijsman2019openssl} repurposed the Diffie-Hellman engine infrastructure, implementing pure QKD in modern cryptographic frameworks is challenging due to these architectural constraints.

Our solution encapsulates both the PQC shared secret and the QKD key identifier into a single KEM operation through an OpenSSL provider. We have implemented two approaches for QKD integration: a client-initiated flow where the client retrieves and transmits the key ID during key generation, and a server-initiated flow similar to García et al. \cite{garcia2023quantum,garcia2024integrating} where the server manages key retrieval. While the server-initiated approach better aligns with TLS 1.3 philosophy, it is only compatible with the stateless ETSI 014 API. The client-initiated approach supports both ETSI 004 and 014 specifications, as ETSI 004 requires sequential session establishment that doesn't naturally fit server-initiated flows.

\subsection{Hybrid QKD-KEM Protocol Flow}

We implemented two distinct approaches for integrating QKD with TLS, each with different tradeoffs regarding protocol alignment and API compatibility. In both approaches, our protocol combines post-quantum security with QKD information-theoretic security, with several consistent design choices: QKD material is always placed last in the shared secret; both parties derive identical shared secrets through concatenation; and the total shared secret length is the sum of the post-quantum secret length and the QKD key size.

\begin{figure*}[tbp]
    \centering
    \includegraphics[width=0.65\textwidth]{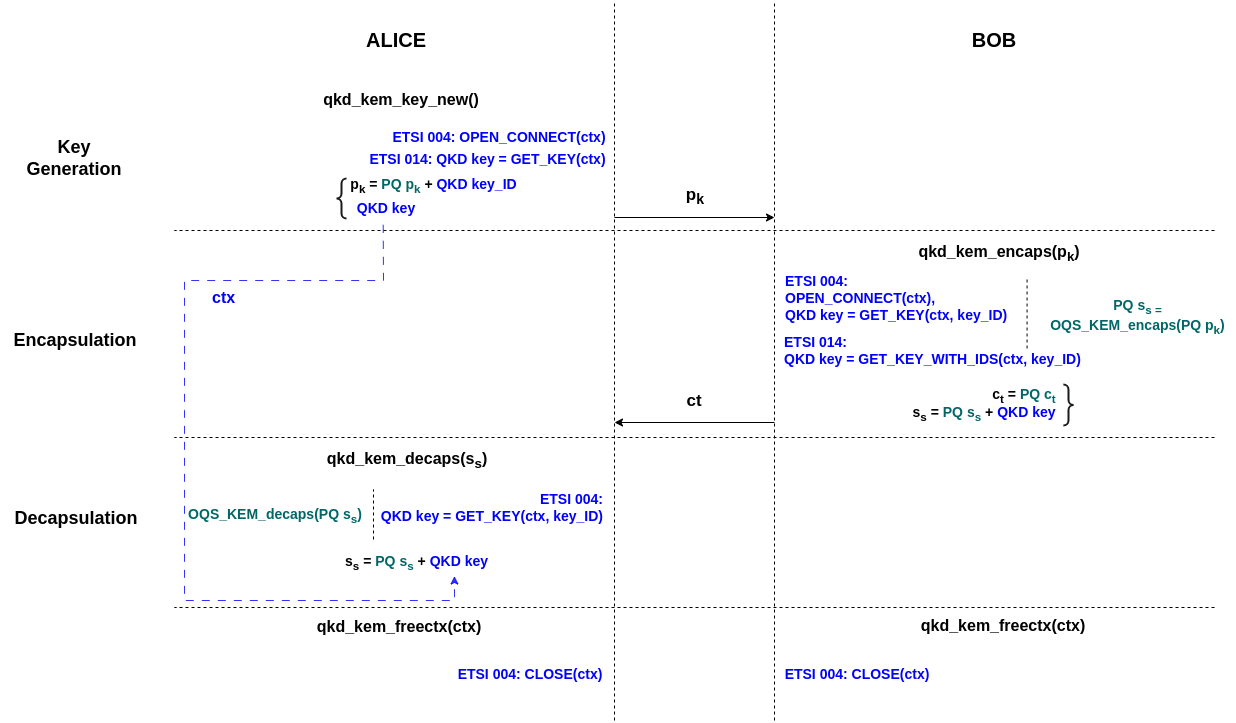}
    \caption{Client-initiated QKD-KEM handshake flow using both ETSI 004 and ETSI 014 APIs. The diagram shows QKD API calls during key generation, encapsulation, and decapsulation. Here, $p_k$ is the public key, $c_t$ is the ciphertext, and $s_s$ is the shared secret; $PQ \ c_t$ and $PQ \ s_s$ denote the post-quantum ciphertext and shared secret, respectively; $ctx$ represents the OpenSSL context. QKD-related API calls and cryptographic/key materials are highlighted in blue, while PQC-related ones are shown in green.}
    \label{fig:qkd_kem_flow}
\end{figure*}

\subsubsection{Client-Initiated Flow}
The client-initiated flow, illustrated in Figure~\ref{fig:qkd_kem_flow}, consists of three phases:

\textbf{Key Generation (Alice):} Alice initiates the protocol by generating the key pair and
retrieving initial QKD material:
\begin{enumerate}
\item Create a provider context and generate a KEM keypair.
\item Initialize the QKD context as initiator.
\item Retrieve the QKD key identifier:
\begin{itemize}
\item ETSI 004: Establish a session via \texttt{OPEN\_CONNECT()}.
\item ETSI 014: Obtain key and ID using \texttt{GET\_KEY()}.
\end{itemize}
\item Transmit the public key and QKD key ID to Bob.
\end{enumerate}
\textbf{Encapsulation (Bob):} Bob performs encapsulation upon receiving Alice's public key and QKD key ID:
\begin{enumerate}
\item Receive Alice's public key and QKD key ID, and invoke the encapsulation operation.
\item Initialize the QKD context as responder.
\item Retrieve the QKD key:
\begin{itemize}
\item ETSI 004: Establish a session via \texttt{OPEN\_CONNECT()} and get the key using \texttt{GET\_KEY(key\_id)}.
\item ETSI 014: Use \texttt{GET\_KEY\_WITH\_IDS()}.
\end{itemize}
\item Perform the PQC encapsulation operation.
\end{enumerate}
\noindent \textbf{Output:} A ciphertext (the PQC part) and a local shared secret obtained by concatenating the PQC secret with the QKD key material.

\textbf{Decapsulation (Alice):} Alice performs decapsulation upon receiving Bob's ciphertext, with QKD key retrieval:
\begin{enumerate}
\item Upon receiving the ciphertext, invoke the decapsulation operation.
\item Retrieve the QKD key (using \texttt{GET\_KEY(key\_id)} for ETSI 004).
\item Recover the PQC shared secret via the decapsulation mechanism.
\end{enumerate}
\noindent \textbf{Output:} The reconstructed shared secret as the concatenation of the PQC and QKD components.

This synchronization is particularly important in the ETSI 004 case, where both parties must establish QKD sessions before retrieving keys. After Alice initiates with \texttt{OPEN\_CONNECT()}, Bob must also call \texttt{OPEN\_CONNECT()} with the received key ID to establish his session before any \texttt{GET\_KEY()} operations can occur. This ensures proper key synchronization and that both parties are ready to perform key retrieval operations. In contrast, ETSI 014's stateless approach means synchronization happens implicitly through the key ID reference system, with Bob's \texttt{GET\_KEY\_WITH\_IDS()} automatically retrieving the correct corresponding key previously obtained by Alice's \texttt{GET\_KEY()}.

\subsubsection{Server-Initiated Flow}
We also implemented a server-initiated approach similar to García et al.~\cite{garcia2023quantum,garcia2024integrating}, where:

\begin{itemize}
    \item \textbf{Client Hello:} The client signals QKD support without retrieving any keys
    \item \textbf{Server Hello:} The server retrieves the QKD key via \texttt{GET\_KEY()} and transmits the key ID to the client
    \item \textbf{Encapsulation:} The client retrieves the corresponding key using \texttt{GET\_KEY\_WITH\_IDS(key\_id)}
\end{itemize}

This server-centric flow better aligns with TLS 1.3 conventions where servers handle more key management responsibilities. However, it is only compatible with ETSI 014 APIs due to their stateless nature. The ETSI 004 specification requires a three-step handshake (initiator calls \texttt{OPEN\_CONNECT()}, ID is communicated, responder calls \texttt{OPEN\_CONNECT()}), making the client-initiated approach necessary for ETSI 004 implementations.

\subsection{OpenSSL Provider Implementation}

\begin{figure}[tbp]
    \centering
    \includegraphics[width=1.0\columnwidth]{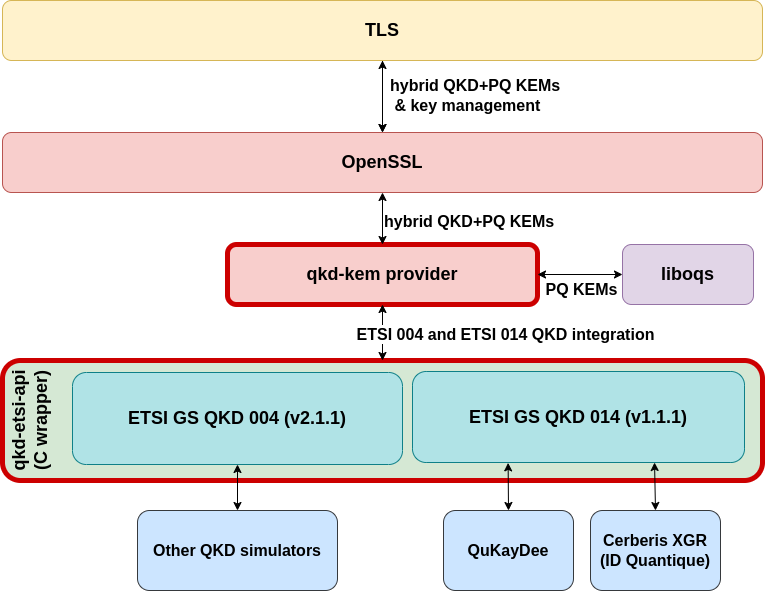}
    \caption{Architecture of the hybrid PQC+QKD integration into TLS via the QKD-KEM Provider.}
    \label{fig:qkd_kem}
\end{figure}

The OQS Provider \cite{oqsprovider} is an Open Quantum Safe provider for OpenSSL (3.x) that enables quantum-safe cryptography by integrating quantum-resistant algorithms into OpenSSL via a single shared library. It relies on liboqs \cite{liboqs}—an open source C library offering implementations and a common API for quantum-safe key encapsulation mechanisms and digital signature schemes. We forked the OQS Provider \cite{qursa2024qkd} (v0.7.0) to integrate QKD support into OpenSSL. A C wrapper library provides interfaces for both ETSI GS QKD 004 V2.1.1 and ETSI GS QKD 014 APIs \cite{etsi2020qkd014}, abstracting hardware dependencies. The ETSI 004 interface provides stream-based key delivery (via \texttt{QKD\_OPEN}, \texttt{QKD\_GET\_KEY}, and \texttt{QKD\_CLOSE}), while the ETSI 014 interface offers block key delivery with unique identifiers (\texttt{GET\_STATUS}, \texttt{GET\_KEY}, and \texttt{GET\_KEY\_WITH\_IDS}). The backend selection (simulated or hardware) is managed at compile time. For development and preliminary testing, we utilized the QuKayDee QKD network simulator \cite{qukaydee2024docs} before transitioning to production hardware.

Our implementation exchanges QKD key identifiers differently depending on the flow: in client-initiated mode, the key ID accompanies the client's public key during key generation, while in server-initiated mode, the server sends the key ID concatenated with the PQC ciphertext during encapsulation. The unified KEM abstraction combines PQC and QKD shared secrets at each endpoint.

We integrate with TLS 1.3's negotiation framework (RFC 8446~\cite{rescorla2018transport}) by registering custom group identifiers (e.g., \texttt{qkd\_kyber768}) transmitted in the client's \texttt{supported\_groups} extension. The \texttt{key\_share} extension carries PQC public keys and, in client-initiated flow, QKD key identifiers. The server selects a supported group in its Server Hello response, ensuring hybrid exchanges occur only when both endpoints support it. This approach offers significant advantages in server-initiated flows, where servers verify client QKD support before retrieving keys, whereas client-initiated flows must retrieve QKD keys before confirming server support, potentially wasting quantum-distributed keys. Network analysis tools such as Wireshark may display these groups as ``Unknown (0x303c)'' as they aren't yet standardized TLS identifiers.

\section{Preliminary Evaluation}
\label{sec:evaluation}

We evaluate our implementation using a benchmarking suite \cite{qursa2024bench} that tests both isolated KEM operations and complete TLS 1.3 handshakes via OpenSSL.

\subsection{Experimental setup}

This section details our proof-of-concept tests for the hybrid QKD-KEM architecture. Tests were conducted on Ubuntu 24.10 (kernel 6.11.0-19-generic) using an AMD Ryzen AI 9 HX 370 processor (24 threads, 12 cores, 4.37GHz) with 32GB RAM. During initial development phases, we leveraged the QuKayDee QKD network simulator for testing before moving to our evaluation using production ID Quantique Cerberis XGR QKD hardware deployed at the University of Vigo, with two nodes connected via dedicated dark fiber functioning as Key Management Entities, while TLS endpoints were located in Madrid. The ETSI 014 API communication uses HTTPS with mutual authentication, though our proof-of-concept transmits QKD keys over HTTPS, creating potential security implications discussed in our Conclusions.

We conducted performance comparisons between our QKD-KEM provider and the standard OQS provider \cite{oqsprovider}, running 50 iterations for each provider and test scenario. The evaluation comprised two phases: isolated key operations (key generation, encapsulation, decapsulation) and full TLS handshakes. The isolated tests utilized \texttt{scripts/run\_qkd\_kem\_bench.sh} from \cite{qursa2024bench}.

For TLS testing, we developed two key components: \texttt{scripts/test\_qkd\_kem\_tls.py}, which implements the core TLS functionality, and \texttt{scripts/run\_tls\_bench.sh}, which automates handshake tests across various KEMs and certificate types. The scripts capture handshake completion times for different combinations of KEMs (including \texttt{mlkem}, \texttt{bike}, \texttt{Frodo} and \texttt{hqc} variants) and certificates (RSA, MLDSA, Falcon). 
Tests employed OpenSSL 3.4.0, liboqs 0.12.0, and oqsprovider 0.8.0. Results are presented in Figures~\ref{fig:oqs_kem_ops_time}, \ref{fig:qkd_kem_ops_time}, and \ref{fig:tls_handshake_time_cerberis}.

\iffalse
\noindent\textcolor{violet}{QKD Hardware:}
\begin{itemize}
    \item \textcolor{violet}{Two nodes Cerberis XGR QKD System \cite{nodes-idq} are connected through a dark fiber that acts as the quantum channel used by the nodes to perform the QKD. Additionally, each node acts as a KME (Key Management Entity) for each extreme of the communication.}
\end{itemize}
\fi

\subsection{Results}

\begin{figure}[tbp]
    \centering
    \includegraphics[width=1.0\columnwidth]{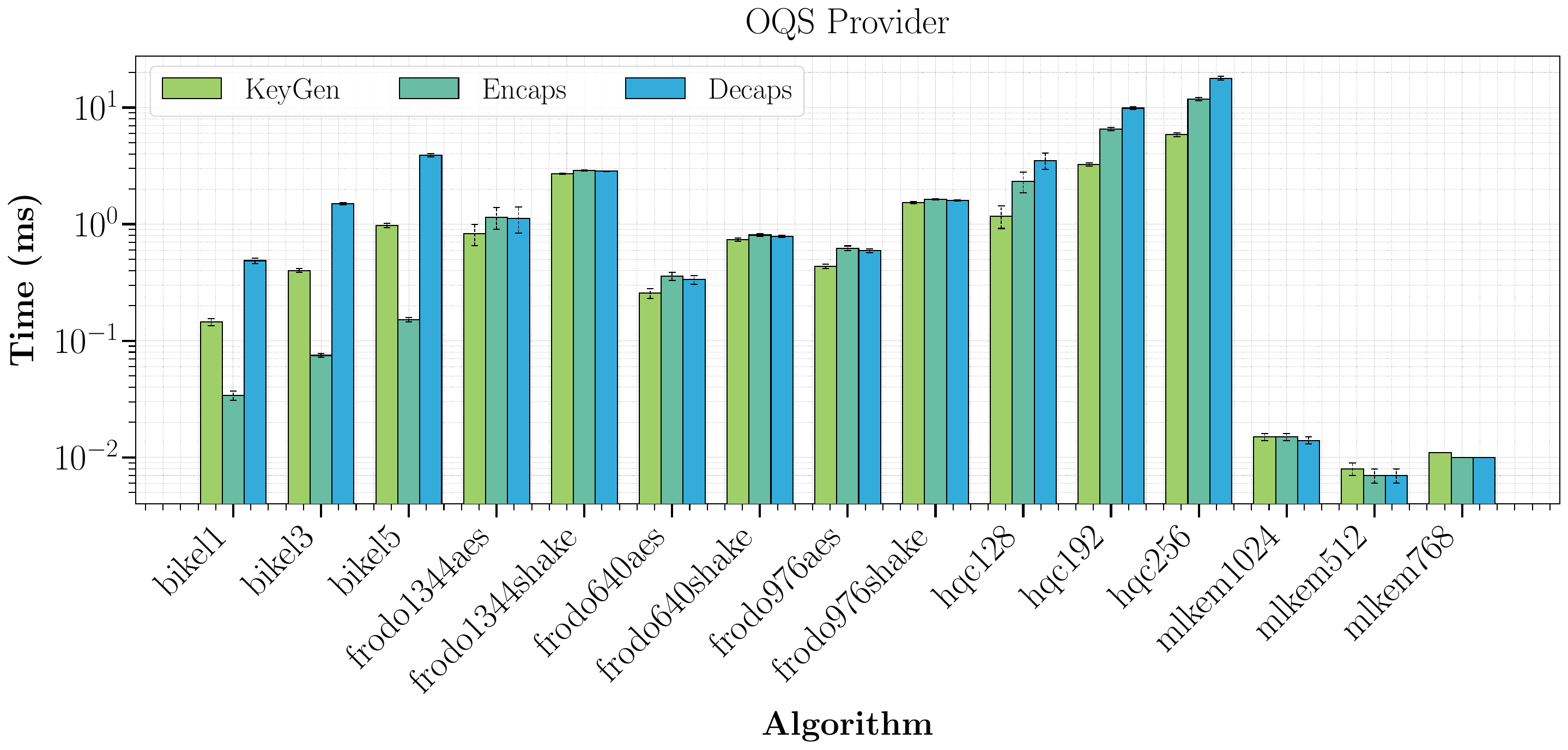}
    \caption{Performance comparison of key generation, encapsulation, and decapsulation operations for different post-quantum KEM algorithms. Times shown in logarithmic scale (milliseconds).}
    \label{fig:oqs_kem_ops_time}
\end{figure}

Our performance evaluation examines three measurements: standalone KEM operations, QKD-KEM operations, and TLS handshakes. For standalone KEMs (Figure \ref{fig:oqs_kem_ops_time}), ML-KEM variants complete in microseconds while HQC algorithms require milliseconds, with key generation being consistently the slowest operation.

\begin{figure}[tbp]
    \centering
    \includegraphics[width=1.0\columnwidth]{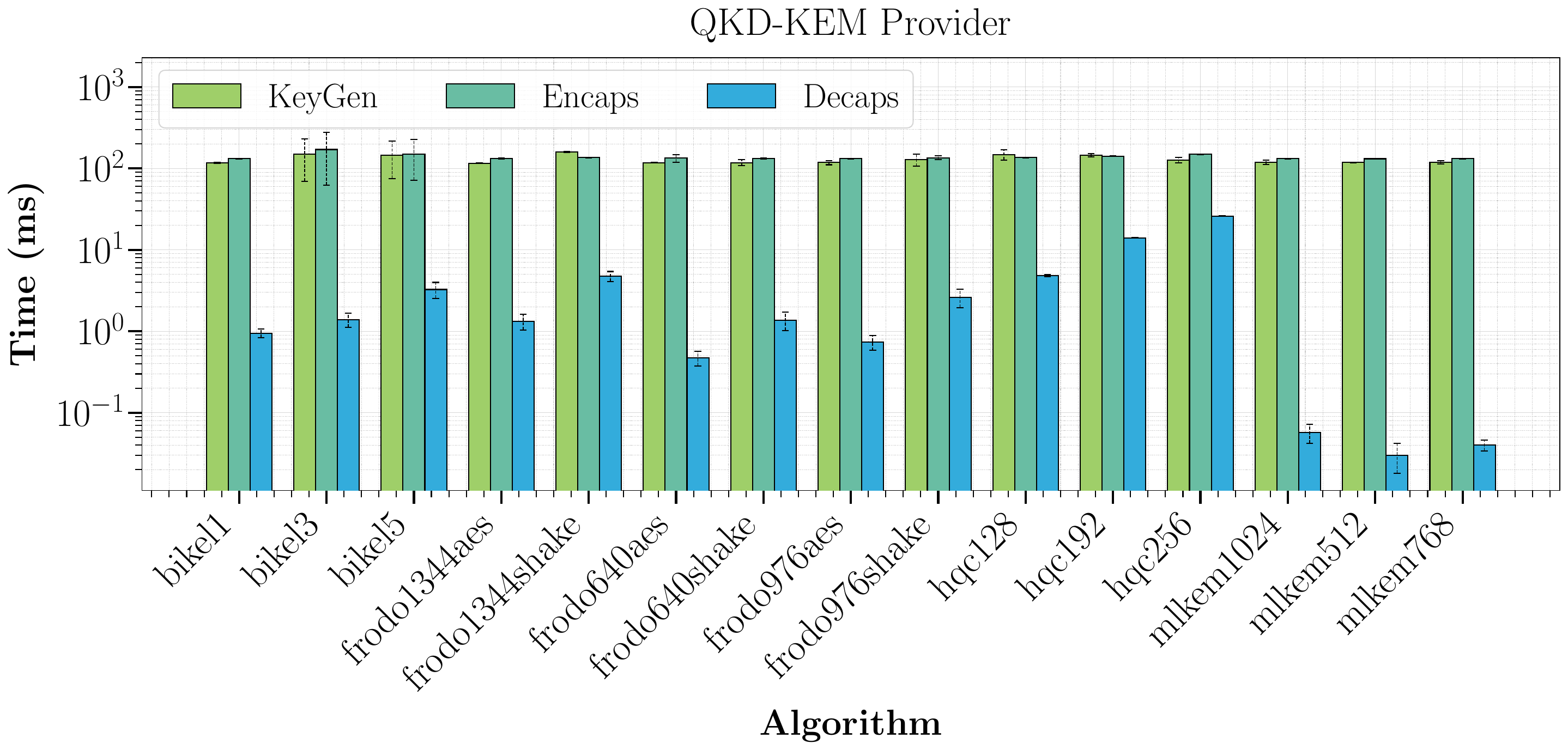}
    \caption{Performance comparison of key generation, encapsulation, and decapsulation operations for hybrid QKD-KEM algorithms. The increased times in key generation and encapsulation reflect the ETSI 014 API calls overhead. Times shown in logarithmic scale (milliseconds).}
    \label{fig:qkd_kem_ops_time}
\end{figure}

Adding QKD functionality (Figure \ref{fig:qkd_kem_ops_time}) introduces a fixed ~100ms overhead in key generation and encapsulation due to ETSI 014 API calls. Decapsulation times remain unchanged as they require no additional QKD API calls. While client-initiated flows place overhead in key generation and encapsulation phases, server-centric approaches (not shown in this paper) shift this load to encapsulation and decapsulation instead.

\iffalse
TLS handshake measurements (Figure \ref{fig:tls_handshake_time}) show standalone PQC handshakes completing in 10-100ms, while hybrid QKD-KEM handshakes take approximately 1000ms. This order-of-magnitude increase stems from QKD key retrieval overhead, which remains constant across PQC algorithms.

\begin{figure}[tbp]
    \centering
    \includegraphics[width=1.0\columnwidth]{tls_kems_comparison_rsa2048_log.pdf}
    \caption{Performance comparison of TLS handshake times using RSA-2048 certificates for standard PQC and hybrid QKD-KEM approaches. The increased times in the hybrid approach reflect the overhead of QKD key retrieval operations. Times shown in logarithmic scale (milliseconds).}
    \label{fig:tls_handshake_time}
\end{figure}
\fi

\begin{figure}[tbp]
    \centering
    \includegraphics[width=1.0\columnwidth]{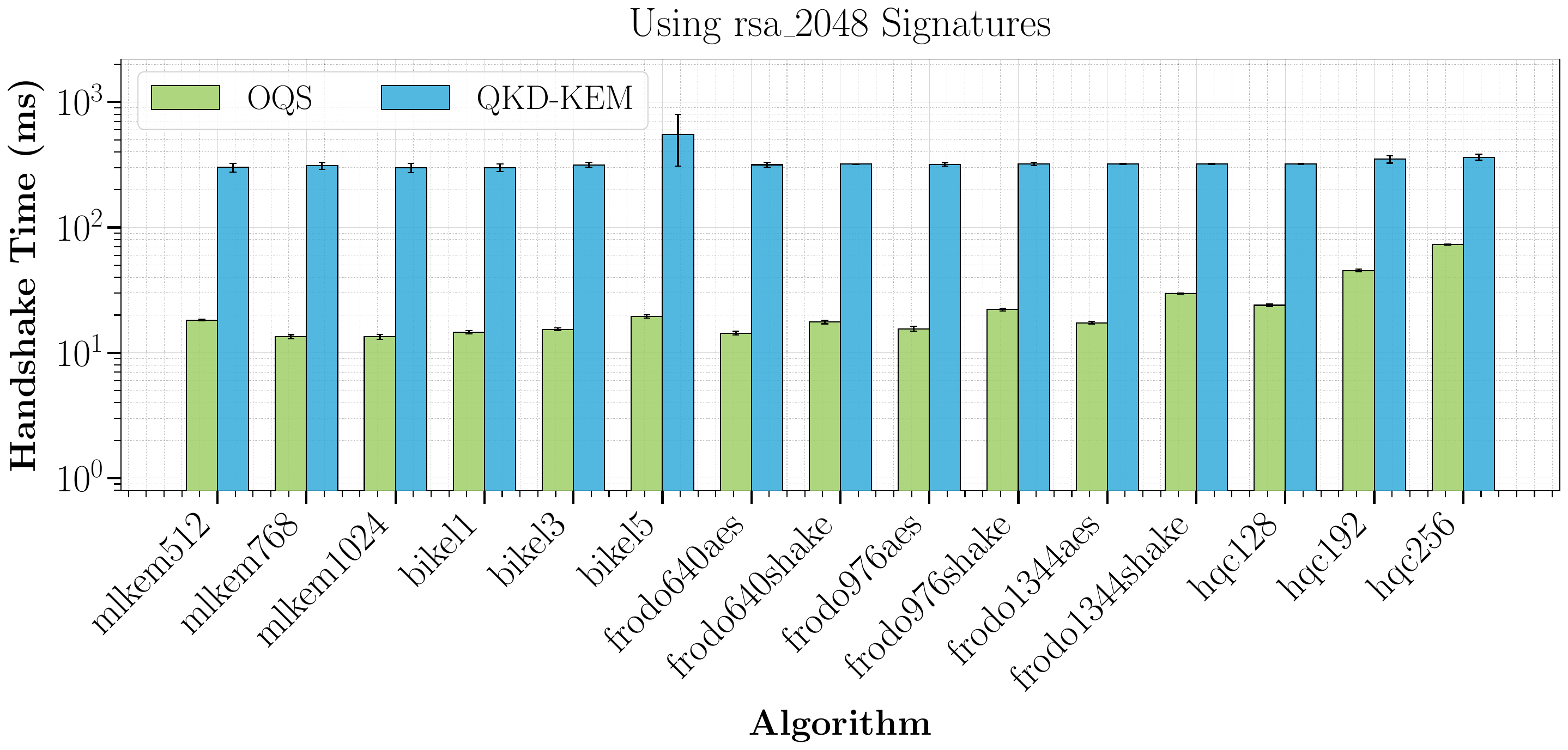}
    \caption{Performance comparison of TLS handshake times using RSA-2048 certificates for standard PQC and hybrid QKD-KEM approaches with production Cerberis XGR QKD hardware. Tests conducted between Madrid (client/server endpoints) and University of Vigo (QKD nodes) show reduced latency compared to simulator-based tests. Times shown in logarithmic scale (milliseconds).}
    \label{fig:tls_handshake_time_cerberis}
\end{figure}

The Cerberis XGR hardware tests (Figure \ref{fig:tls_handshake_time_cerberis} and Table \ref{tab:tls_comparison}) show complete handshake times of 300-350 ms for most algorithms, which remain practical for real-world applications despite the overhead compared to pure PQC approaches. The observed performance is influenced by network conditions between Madrid and Vigo, as well as the ETSI API implementation of the commercial hardware.

It's important to note that in a production environment with proper QKD network architecture and co-located nodes, we would expect significantly lower latencies for the QKD operations.

\begin{table}[htbp]
  \centering
  \caption{TLS Handshake Performance Comparison Between OQS and QKD-KEM Providers for 20 iterations}
  \label{tab:tls_comparison}
  \begin{tabular}{|l|r|r|r|r|}
    \hline
    \textbf{Algorithm} & \multicolumn{2}{c|}{\textbf{OQS (ms)}} & \multicolumn{2}{c|}{\textbf{QKD (ms)}} \\
    \cline{2-5}
    & \textbf{Mean} & \textbf{Std} & \textbf{Mean} & \textbf{Std} \\
    \hline
    mlkem512 & 18.22 & 0.26 & 300.96 & 23.67 \\
    mlkem768 & 13.47 & 0.48 & 310.76 & 20.03 \\
    mlkem1024 & 13.43 & 0.58 & 298.50 & 25.17 \\
    \hline
    bikel1 & 14.61 & 0.39 & 300.10 & 20.95 \\
    bikel3 & 15.40 & 0.39 & 315.78 & 14.90 \\
    bikel5 & 19.49 & 0.55 & 551.01 & 243.05 \\
    \hline
    frodo640aes & 14.35 & 0.47 & 316.01 & 13.57 \\
    frodo640shake & 17.60 & 0.57 & 320.42 & 1.57 \\
    frodo976aes & 15.53 & 0.71 & 318.44 & 10.55 \\
    frodo976shake & 22.19 & 0.55 & 322.12 & 9.58 \\
    frodo1344aes & 17.37 & 0.42 & 320.69 & 1.94 \\
    frodo1344shake & 29.70 & 0.35 & 321.77 & 3.81 \\
    \hline
    hqc128 & 23.93 & 0.58 & 320.57 & 2.14 \\
    hqc192 & 45.35 & 1.07 & 349.89 & 23.99 \\
    hqc256 & 72.92 & 0.48 & 362.40 & 19.96 \\
    \hline
  \end{tabular}
\end{table}

\section{Discussion}\label{sec:discussion}

Our proof-of-concept demonstrates the feasibility of QKD-PQC integration in TLS, achieving handshake times under 1 second with the remote QKD node setup. However, our implementation, where QKD keys traverse HTTPS connections, lacks true quantum security—production environments require endpoints co-located with QKD nodes within secure network segments to prevent key exposure and reduce latency.

The server-initiated approach aligns with TLS 1.3's design philosophy, where servers control cryptographic parameters, key shares, PSK selection, and session tickets. This server-centric pattern complements QKD integration while preventing unnecessary key consumption—an advantage where QKD keys are limited resources requiring specialized quantum channels. The simpler client-initiated approach risks wasting QKD resources when servers lack QKD support. However, despite these advantages, the client-initiated approach remains necessary for ETSI 004 integration due to its stateful nature and sequential session establishment requirements that don't naturally fit the server-initiated flow model.

Our hybrid approach provides enhanced security through complementary protection: PQC algorithms defend against quantum computing attacks, while QKD's information-theoretic security protects against cryptanalytic breakthroughs. This redundancy requires an adversary to compromise both components to breach the connection. Importantly, the cost of hybridizing with PQC is small and often of the same order as QKD operations, making it a cost-effective way to provide an additional security layer. This directly addresses the concerns of cybersecurity agencies~\cite{QKD2024} regarding QKD's current maturity limitations in practical deployments.

Further optimizations could include parallelizing ETSI API requests with encapsulation operations and implementing key pre-fetching mechanisms to reduce handshake times for production deployments.

\section{Conclusions and Future Work}\label{sec:conclusions}

We presented a hybrid QKD-KEM protocol integrating QKD with post-quantum cryptography into TLS via OpenSSL's provider interface. Our approach demonstrates the feasibility of combining quantum and post-quantum methods through adaptation of KEM primitives. However, a new provider API specifically tailored for QKD exchanges could better accommodate quantum key establishment while maintaining interoperability with classical and post-quantum operations.

Our future work will focus on:
\begin{itemize} 
    \item \textbf{Hybrid Schemes:} Developing triple hybrid schemes integrating QKD, PQC, and traditional cryptography.
    \item \textbf{Secure Network Architecture Testing:} Testing with QKD hardware where endpoints are co-located with QKD nodes within secure network boundaries, ensuring keys never traverse quantum-insecure networks and validating our approach in real-world conditions.
    \item \textbf{IPSec IKEv2 Integration:} Extending our integration to IPSec IKEv2 using the strongSwan plugin interface \cite{strongswan-github} to demonstrate broader applicability across secure communication protocols.
    \item \textbf{ETSI 004 API Testing:} Evaluating compatibility and performance with industry-standard QKD key delivery protocols.
\end{itemize}

\section*{Acknowledgments}
This work was supported under the grant TED–2021–130369B–C32 funded by MICIU/AEI/
10.13039/501100011033 and by the “European Union NextGenerationEU/PRTR” and the grant  PID2020–113795RB–C32 funded by MICIU/AEI/
10.13039/501100011033. In addition, it was partially supported by the I-Shaper Strategic Project (C114/23), due to the collaboration agreement signed between the Instituto Nacional de Ciberseguridad (INCIBE) and the UC3M; this initiative is being carried out within the framework of the Recovery, Transformation and Resilience Plan funds, funded by the European Union (Next Generation).

\bibliographystyle{unsrt}
\bibliography{main}

\begin{thebibliography}{10}

\bibitem{rescorla2018transport}
Eric Rescorla.
\newblock The transport layer security (tls) protocol version 1.3, 2018.

\bibitem{gisin2002quantum}
Nicolas Gisin, Gr{\'e}goire Ribordy, Wolfgang Tittel, and Hugo Zbinden.
\newblock Quantum cryptography.
\newblock {\em Reviews of modern physics}, 74(1):145, 2002.

\bibitem{QKD2024}
French Cybersecurity~Agency (ANSSI), Federal~Office for Information Security~(BSI), Netherlands National Communications Security~Agency (NLNCSA), and Swedish Armed~Forces Swedish National Communications Security~Authority.
\newblock Position paper on quantum key distribution, January 2024.

\bibitem{bernstein2017post}
Daniel~J Bernstein and Tanja Lange.
\newblock Post-quantum cryptography.
\newblock {\em Nature}, 549(7671):188--194, 2017.

\bibitem{stebila2020hybrid}
Douglas Stebila, Scott Fluhrer, and Shay Gueron.
\newblock Hybrid key exchange in tls 1.3.
\newblock {\em IETF draft}, 2020.

\bibitem{sfaxi2005using}
MA~Sfaxi, S~Ghernaouti-H{\'e}lie, G~Ribordy, and O~Gay.
\newblock Using quantum key distribution within ipsec to secure man communications.
\newblock {\em Proceedings of metropolitan area networks (man2005)}, 2005.

\bibitem{berzanskis2009method}
Audrius Berzanskis, Harri Hakkarainen, Keun Lee, and Muhammad~Raghib Hussain.
\newblock Method of integrating qkd with ipsec, October 2009.

\bibitem{aguado2017hybrid}
Alejandro Aguado, Victor Lopez, Jesus Martinez-Mateo, Thomas Szyrkowiec, Achim Autenrieth, Momtchil Peev, Diego Lopez, and Vicente Martin.
\newblock Hybrid conventional and quantum security for software defined and virtualized networks.
\newblock {\em Journal of Optical Communications and Networking}, 9(10):819--825, 2017.

\bibitem{rijsman2019openssl}
Bruno Rijsman, Yvo Keuter, and Tim Janssen.
\newblock Integration of quantum key distribution in openssl.
\newblock In {\em Pan-European Quantum Internet Hackathon}, Delft, November 2019. RIPE Labs.
\newblock Developed at QuTech, TU Delft.

\bibitem{dowling2020many}
Benjamin Dowling, Torben~Brandt Hansen, and Kenneth~G Paterson.
\newblock Many a mickle makes a muckle: A framework for provably quantum-secure hybrid key exchange.
\newblock In {\em International Conference on Post-Quantum Cryptography}, pages 483--502. Springer, 2020.

\bibitem{huang2020practical}
Leilei Huang, Kai Feng, and Chongjin Xie.
\newblock A practical hybrid quantum-safe cryptographic scheme between data centers.
\newblock In {\em Emerging Imaging and Sensing Technologies for Security and Defence V; and Advanced Manufacturing Technologies for Micro-and Nanosystems in Security and Defence III}, volume 11540, pages 30--35. SPIE, 2020.

\bibitem{dervisevic2021overview}
Emir Dervisevic and Miralem Mehic.
\newblock Overview of quantum key distribution technique within ipsec architecture.
\newblock {\em arXiv preprint arXiv:2112.13105}, 2021.

\bibitem{kozlovivcs2023quantum}
Sergejs Kozlovi{\v{c}}s, Kri{\v{s}}j{\=a}nis Petru{\v{c}}e{\c{n}}a, D{\=a}vis L{\=a}ri{\c{n}}{\v{s}}, and Juris V{\=\i}ksna.
\newblock Quantum key distribution as a service and its injection into tls.
\newblock In {\em International Conference on Information Security Practice and Experience}, pages 527--545. Springer, 2023.

\bibitem{garcia2023enhancing}
Carlos~Rubio Garc{\'\i}a, Simon Rommel, Juan Jose~Vegas Olmos, and Idelfonso~Tafur Monroy.
\newblock Enhancing the security of software defined networks via quantum key distribution and post-quantum cryptography.
\newblock In {\em International Symposium on Distributed Computing and Artificial Intelligence}, pages 428--437. Springer, 2023.

\bibitem{garcia2023quantum}
Carlos~Rubio Garcia, Abraham~Cano Aguilera, Juan Jose~Vegas Olmos, Idelfonso~Tafur Monroy, and Simon Rommel.
\newblock Quantum-resistant tls 1.3: A hybrid solution combining classical, quantum and post-quantum cryptography.
\newblock In {\em 2023 IEEE 28th International Workshop on Computer Aided Modeling and Design of Communication Links and Networks (CAMAD)}, pages 246--251. IEEE, 2023.

\bibitem{garcia2024integrating}
Carlos~Rubio Garcia, Abraham Cano, JJ~Vegas Olmos, Simon Rommel, and Idelfonso~Tafur Monroy.
\newblock Integrating quantum key distribution into tls 1.3: A transport layer approach to quantum-resistant communications in optical networks.
\newblock In {\em Optical Fiber Communication Conference}, pages Th3B--3. Optica Publishing Group, 2024.

\bibitem{rencis2024hybrid}
Edgars Rencis, Juris V{\=\i}ksna, Sergejs Kozlovi{\v{c}}s, Edgars Celms, D{\=a}vis~J{\=a}nis L{\=a}ri{\c{n}}{\v{s}}, and Kri{\v{s}}j{\=a}nis Petru{\v{c}}e{\c{n}}a.
\newblock Hybrid qkd-based framework for secure enterprise communication system.
\newblock {\em Procedia Computer Science}, 239:420--428, 2024.

\bibitem{hoque2024exploring}
Sanzida Hoque, Abdullah Aydeger, and Engin Zeydan.
\newblock Exploring post quantum cryptography with quantum key distribution for sustainable mobile network architecture design.
\newblock In {\em Proceedings of the 4th Workshop on Performance and Energy Efficiency in Concurrent and Distributed Systems}, pages 9--16, 2024.

\bibitem{itu2024qkdtls}
{ITU-T}.
\newblock {Draft Recommendation ITU-T Y.QKD-TLS: Quantum Key Distribution integration with Transport Layer Security 1.3}.
\newblock Draft Recommendation SG13-TD412/WP3, International Telecommunication Union, 3 2024.
\newblock Expected Q3 2024.

\bibitem{alia2024100}
Obada Alia, Albert Huang, Huan Luo, Omar Amer, Marco Pistoia, and Charles Lim.
\newblock 100 gbps quantum-safe ipsec vpn tunnels over 46 km deployed fiber.
\newblock {\em arXiv preprint arXiv:2405.04415}, 2024.

\bibitem{qursa2024qkd}
QURSA Project.
\newblock Qkd-kem provider, 2024.
\newblock OpenSSL 3.0 provider implementing hybrid QKD-PQC key encapsulation.

\bibitem{blanco2024qkd}
QURSA Project.
\newblock Qkd etsi api implementation, 2024.
\newblock Implementation of ETSI GS QKD 004 V2.1.1 API specification.

\bibitem{qursa2024bench}
QURSA Project.
\newblock Qkd-kem benchmarking suite, 2024.
\newblock Performance evaluation framework for hybrid QKD-KEM operations.

\bibitem{openssl2024keyexch}
{OpenSSL Project}.
\newblock Openssl provider keyexch documentation.
\newblock \url{https://docs.openssl.org/3.3/man7/provider-keyexch}, 2024.
\newblock OpenSSL Documentation version 3.3, accessed March 9, 2025.

\bibitem{openssl2024kem}
{OpenSSL Project}.
\newblock Openssl provider kem documentation.
\newblock \url{https://docs.openssl.org/3.1/man7/provider-kem}, 2024.
\newblock OpenSSL Documentation version 3.1, accessed March 9, 2025.

\bibitem{oqsprovider}
Michael Baentsch, Christian Paquin, Richard Levitte, Basil Hess, Julian Segeth, Alex Zaslavsky, Will Childs-Klein, Thomas Bailleux, Felipe Ventura, and {Open Quantum Safe Project}.
\newblock Oqs provider: Open quantum safe provider for {OpenSSL} 3.x, 2024.
\newblock OpenSSL 3.x provider implementing post-quantum cryptographic algorithms for key establishment and signatures.

\bibitem{liboqs}
{Open Quantum Safe Project}.
\newblock liboqs: {C} library for quantum-safe cryptographic algorithms, 2024.
\newblock Open source C library implementing quantum-safe key encapsulation mechanisms and digital signature algorithms.

\bibitem{etsi2020qkd014}
ETSI.
\newblock Etsi gs qkd 014 v1.1.1: Quantum key distribution (qkd); protocol and data format of rest-based key delivery api.
\newblock Technical Specification GS QKD 014 V1.1.1, European Telecommunications Standards Institute, 2020.

\bibitem{qukaydee2024docs}
QuKayDee Team.
\newblock Qukaydee: Qkd netwfork simulator documentation.
\newblock \url{https://qukaydee.com/pages/about}, 2024.
\newblock Online documentation, version f135283, accessed February 3, 2025.

\bibitem{strongswan-github}
{strongSwan Project}.
\newblock strongswan: {IPsec}-based {VPN} solution, 2024.
\newblock Open Source Project Repository.

\end{thebibliography}

\end{document}